\documentclass[twocolumn]{aastex63}

\usepackage{graphicx}
\usepackage{lipsum}
\usepackage{wrapfig}
\usepackage{rotating}
\usepackage[version=3]{mhchem} 

\def\invcm{cm$^{-1}$~}

\shortauthors{Jorge, Wordsworth and Adams}

\begin{document}

\title{Greenhouse warming potential of a suite of gas species on early Mars evaluated using a radiative-convective climate model}

\author{Jason Jorge}
\affil{Department of Earth and Planetary Sciences, Harvard University, Cambridge, MA, USA}
\author{Robin Wordsworth}
\affil{Department of Earth and Planetary Sciences, Harvard University, Cambridge, MA, USA}
\affil{School of Engineering and Applied Sciences, Harvard University, Cambridge, MA, USA}
\author{Danica Adams}
\affil{Department of Earth and Planetary Sciences, Harvard University, Cambridge, MA, USA}

\begin{abstract}
Abundant geomorphological and geochemical evidence of liquid water on the surface of early Mars during the late Noachian and early Hesperian periods needs to be reconciled with a fainter young Sun. While a dense \ce{CO2} atmosphere and related warming mechanisms are potential solutions to the early Mars climate problem, further investigation is warranted. Here, we complete a comprehensive survey of the warming potential of all known greenhouse gases and perform detailed calculations for 15 different minor gas species under early Martian conditions. We find that of these 15 species, \ce{H2O2}, \ce{HNO3}, \ce{NH3}, \ce{SO2}, and \ce{C2H4} cause significant greenhouse warming at concentrations of $\sim$0.1~ppmv or greater. However, the most highly effective greenhouse gas species also tend to be {more condensable, soluble and} vulnerable to photolytic destruction. To provide a reference for future atmospheric evolution and photochemical studies, we have made our warming potential database freely available online.
\end{abstract}

\section{Plain Text Summary}
Several billion years ago, geological evidence indicates that Mars was warmer and wetter, with flowing rivers and lakes on its surface. This is hard to explain because the Sun was fainter then, and Mars's orbit is more distant than Earth's. The atmosphere of early Mars was probably thicker, but additional greenhouse gases besides carbon dioxide are needed to provide enough warming. Here we perform a survey of the warming effect of all plausible greenhouse gases on early Mars. We isolate a few species that are particularly effective at warming the planet, calculate how much they raise temperatures at different concentrations, and discuss whether they could have built up to high concentrations in Mars's early atmosphere. Our work provides an important reference for future calculations that incorporate chemistry and other effects.

\section{Introduction}

Mars today is a cold and dry planet, with mean surface temperatures of around 210 K and a \ce{CO2}-dominated atmosphere less than 1$\%$ the volume of Earth's \citep{Martinez2017-nb}. However, extensive geomorphological and geochemical evidence indicate that surface conditions were very different in its early history. Valley networks \citep{Di_Achille2010-kf,Hynek2010-bp}, open-basin lakes \citep{Fassett2008-lp}, and fluvial conglomerates \citep{Williams2013-pd} point to episodically warmer and wetter conditions, with particularly strong evidence for extensive surface alteration by liquid water seen between $\sim$3 and 3.8 Ga, based on crater counting chronology \citep{Werner2011-qr}. 

Mars's distant orbit and the faintness of the young Sun mean that surface temperatures would have been well below zero in the absence of greenhouse warming from a thicker early atmosphere \citep{Haberle2017-px,Wordsworth2016-ua}. While a thicker \ce{CO2} atmosphere can help solve the early martian climate problem, it does not appear sufficient alone \citep{Forget2013-zd,Wordsworth2013-ks,Ramirez2014-vt,kerber2015sulfur,kamada2021global}.  {Other dominant atmospheric constituents on early Mars are possible, such as \ce{H2} or \ce{N2}, but are typically considered less likely. } 

Other solutions such as impact-induced steam atmospheres and \ce{SO2}-induced warming from volcanism have been proposed but suffer shortcomings that make them unlikely as primary explanations \citep{Segura2002-hp,Halevy2007-vu,Segura2012-fl,Halevy2014-jc}. 
{Aerosols, including dust, volcanic sulfuric acid droplets and \ce{H2O} and \ce{CO2} clouds have a variety of effects depending on their location and microphysical properties \citep{Forget2013-zd,Wordsworth2013-ks,Urata2013-lw,kerber2015sulfur,kite2021warm,steakley2023impact}. }
Finally, reducing gases, particularly \ce{H2}, can cause strong warming in combination with \ce{CO2} due to collision-induced absorption effects \citep{Ramirez2014-vt,Wordsworth2017-kc,Turbet2019-ko,steakley2023impact}. This appears to be a promising solution to the faint young Sun problem, but questions remain about how abundant such gases may have been in the martian atmosphere through time, motivating further study.

There has not yet been a comprehensive survey of the warming potential of all known greenhouse gases on early Mars. Such an objective is important, because it can motivate future investigations that couple climate modeling to atmospheric chemical and study of sources and sinks. Here we survey the greenhouse warming potential of the entire HITRAN database \citep{Gordon2022-mt}, and perform detailed calculations for 15 different minor gas species in the early Martian atmosphere. Computing the radiative potential of each of these gases across a wide range of mixing ratios under early Martian conditions, we provide a database of greenhouse warming for early Mars. In doing so, we expand the parameter space of possible warming mechanisms, and highlight gases for further investigation.

In Section~\ref{sec:methods}, we discuss the one-dimensional radiative-convective model used to compute surface temperature and its validation, the selection of gas species, and our choices of uncertain parameters such as \ce{CO2} line broadening coefficients. In Section 3, we discuss the results of our radiative calculations for all gases at 1 bar and the strongest absorbers at different \ce{CO2} pressures. Finally, we discuss the possible sources and sinks of our strongest absorbers in Section 4. 

\section{Methods}\label{sec:methods}
\subsection{Model Description}

We adapt the one-dimensional radiative-convective model PCM LBL \citep{Wordsworth2021-qg} to determine the increase in surface temperature due to the greenhouse effect of minor gas species on early Mars. This iterative line-by-line spectral code allows us to perform high accuracy globally averaged calculations across a wide range of atmospheric compositions. 

PCM LBL calculates surface and atmospheric temperatures in equilibrium for a given solar flux, atmospheric composition and surface properties. Our model atmosphere is vertically divided into 50 layers from the surface to a minimum pressure of 1 Pa, and simulations were run for 300 time steps. Line data from the HITRAN2020 molecular spectroscopic database \citep{Gordon2022-mt} are used to compute absorption cross sections at wavenumber points on a temperature-log pressure grid of 12 interpolation points in 15 K intervals for each absorbing species. The vertical grid is set so that logarithms of pressure are evenly spaced, with their values determined by the initial conditions of surface temperature and surface pressure \citep{Ding2019-uf}. As the model iterates toward the radiative-convective equilibrium state and the atmospheric temperature profile shifts, new absorption cross sections are interpolated from this grid to calculate the adjusted monochromatic optical depth and radiative fluxes. To calculate absorption cross sections, line strengths are scaled from their reference values using the standard HITRAN formula \citep{Gordon2022-mt}
\begin{equation}
    S_{ij}(T) = S_{ij}(T_0) \frac{Q(T_0)}{Q(T)} \frac{1-e^{-h\nu_{ij}/k_BT}}{1-e^{-h\nu_{ij}/k_BT_0}} \frac{e^{-h\nu_i/k_BT}}{e^{-h\nu_i/k_BT_0}}
    \label{eq:linestrength}
\end{equation}
where \emph{i} and \emph{j} represent the ground and excited states of the transition, respectively; \emph{$S_{ij}$} is the line strength; \emph{Q} is  the total internal partition sum (TIPS); \emph{$\nu_{ij}$} is the line location; \emph{$\nu_i$} is the ground state frequency; \emph{T} is the temperature; \emph{$T_0$} = 296 K is the reference temperature; \emph{h} is the Planck constant; and \emph{$k_B$} is the Boltzmann constant. Line broadening due to both collisional and Doppler effects are taken into account by representing lineshape as a Voigt profile via the Humlíček algorithm \citep{Humlicek1982-zw,Wordsworth2017-kc}. Line position corrections due to pressure shifts are also included where HITRAN data is available \citep{Gordon2022-mt}.

\ce{CO2} is the bulk atmospheric constituent in all of our simulations. 
\ce{CO2} pressures are varied between 0.5 and 2 bar based on estimates of the possible range of pressures during the late Noachian to early Hesperian periods \citep{Hu2015-fc,Jakosky2019-rf}. \ce{CO2} spectral lines are truncated at 500~\invcm to include sub-Lorentzian behavior of spectral lines far from the line center; for other gas species, spectral lines are truncated at 25~\invcm and pure Lorentzian behavior is assumed. We use 2000 wavenumber points for both shortwave and longwave calculations. Increases in spectral resolution did not strongly affect spectrally integrated flux and heating rates in the lower atmosphere or equilibrium surface temperature. To improve efficiency of the line-by-line calculation, we also removed weak lines from the HITRAN2020 data set, which we defined as lines with strengths below $1\times 10^{-30}$~\invcm~/(molecule cm$^{-2}$) at 1000 K. \ce{CO2} collision-induced absorption (CIA) is calculated using the Gruszka-Borysow-Baranov (GBB) parameterization \citep{Gruszka1998-vk,Baranov2004-xp,Wordsworth2010-wu}. Eight quadrature points total are used for shortwave and longwave radiative flux calculations, four upwelling and four downwelling. This eight-stream radiative calculation is computed at the pure-absorption limit. 

The effects of Rayleigh scattering by \ce{CO2} are included by increasing the reflectivity of the surface rather than the top of the atmosphere. This works because molecular absorption is only important in the thermal to near-IR for the atmospheres we are studying, while Rayleigh scattering occurs in visible and shorter wavelengths \citep{Ding2019-uf}. The most recent version of PCM LBL can perform multiple scattering calculations \citep{ding2022prospects}, but the scattering effects of clouds and aerosols are neglected here for simplicity. The surface albedo before Rayleigh adjustment is applied is set to $0.2$, a value appropriate for the present-day martian surface.{Surface albedo is uncertain for early Mars. A basaltic unoxidized rock surface would have lower albedo, as would any regions of surface ocean unobscured by clouds. Conversely, clouds or oxidized dust in the atmosphere could raise albedo. However, global warming potential of a given greenhouse gas is not strongly dependent on our choice here.} 

To compute surface temperature, outgoing longwave radiation (OLR) is first calculated by integrating the monochromatic equation for upwelling radiative flux \citep{Schaefer2016-rg}. The difference between OLR and absorbed shortwave radiation (ASR) is then used to iterate the model towards a steady-state surface temperature and atmospheric temperature profile, following \cite{Wordsworth2017-kc}. Here ASR is defined as 
\begin{equation}
ASR = \frac{(1-A_p)F_{sol}}{4},
\end{equation}
where $A_p$ is the planetary albedo and $F_{sol}$ is the solar flux. The solar flux is taken to be $F_{sol} = 0.75 \times 590~$W~m$^{-2}$ where 590~W~m$^{-2}$ is the Earth's solar constant scaled by Mars' average orbital distance and 0.75 is the approximate effect of the fainter early Sun during the late Noachian period. 

\ce{H2O} abundance at each layer is calculated using the saturation vapor pressure curve and a tropospheric relative humidity value of \emph{$\alpha_{sat}$} = 0.8 and a heat capacity of the atmosphere of \emph{$c_p$} = 800~J~kg$^{-1}$~K$^{-1}$. When the profile becomes unstable to moist convection, it relaxes to a moist adiabatic profile. The atmospheric temperature profile is assumed to be a moist adiabat of \ce{H2O} or \ce{CO2} until sub-saturation occurs, at which point the temperature profile relaxes to simple radiative equilibrium.

{We tested for condensation of greenhouse gases by incorporating saturation vapor curves for minor species in PCM LBL. These curves were modeled using the Antoine equation using data from \citep{yaws2015yaws}. We found that for the range of conditions we considered, condensation was important for two species: \ce{H2O2} and \ce{HNO3}. } Finally, for intercomparison of the output of PCM LBL with other 1D radiative-convective models such as Clima \citep{Kasting1988-wr,Kopparapu2013-ig} and the linearized flux evolution (LiFE) model developed by \cite{,Robinson2018-gt}, see \cite{Ding2019-uf}.

\subsection{Gas Selection}

PCM LBL was used to calculate the greenhouse warming potential of \ce{O3}, \ce{N2O}, \ce{CO}, \ce{CH4}, \ce{SO2}, \ce{H2S}, \ce{NO2}, \ce{NH3}, \ce{HNO3}, \ce{HBr}, \ce{OCS}, \ce{H2CO}, \ce{HCN}, \ce{H2O2}, \ce{C2H6}, and \ce{C2H4}. These were selected from the list of all available gases with line-by-line data in the HITRAN2020 database  (55 total). Many gases were excluded because they are unlikely to accumulate in any meaningful amount to act as an important greenhouse gas in the early Martian atmosphere. Radical species such as $\textrm{OH}$ were excluded, for example, because they are highly reactive and very unlikely to accumulate to climatically important concentrations in the atmosphere. Other gases were excluded because of their low infrared absorptivity. For example, homonuclear diatomic molecules such as \ce{O2} and \ce{N2} were excluded because their electric dipole moment is zero and the molecules are IR-inactive, and collision-induced absorption is not important when the gas is a minor species.

\subsection{\ce{CO2} Line Broadening}
\label{line_broaden}
\begin{figure}
    \centering
    \includegraphics[width=0.5\textwidth]{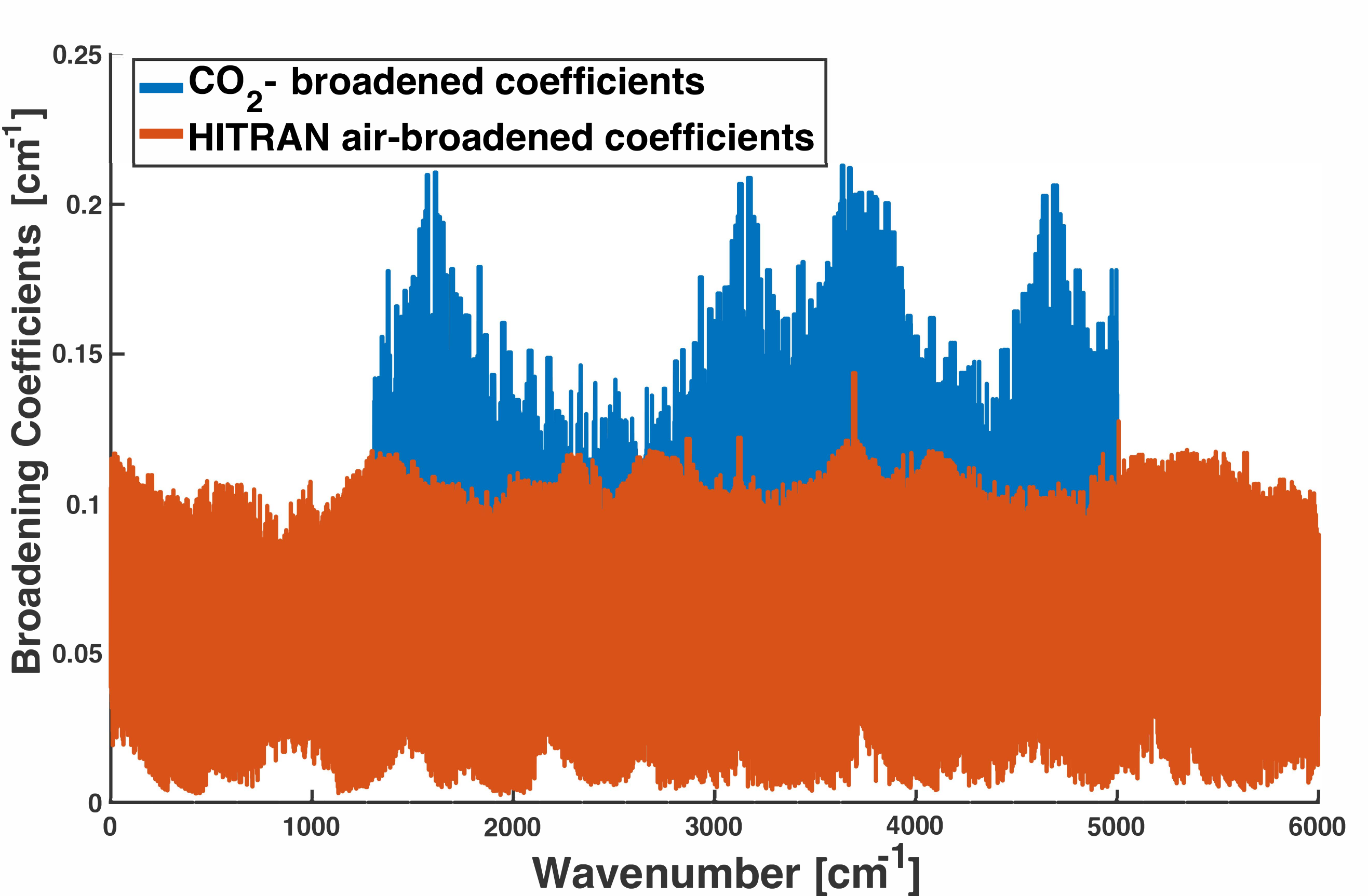}
    \caption{A comparison between \ce{CO2}-broadening coefficients \citep{Regalia2019-tn} and air-broadening coefficients \citep{Gordon2022-mt} for \ce{H2O}. \ce{CO2}-broadened coefficients appear to differ from air-broadened coefficients by a mean factor of $\sim$1.6, in agreement with a mean factor of 1.58 determined by \cite{Deichuli2022-sp}.}
    \label{fig:h2o_gamma}
\end{figure}

Line-shape parameters such as air- and self-broadening coefficients (\emph{$\gamma_L$} and \emph{$\gamma_{self}$} respectively), the temperature exponent of air-broadening (\emph{$n_{air}$}), and the air-pressure-induced line shift (\emph{$\delta_{air}$}) differ from one gas to another. While spectroscopic data for Earth-like (\ce{N2}-dominated) atmospheres are abundant and well-constrained, estimates of broadening parameters for \ce{CO2}-dominated atmospheres have only become available for a few species in recent years. The HITRAN2020 database contains pressure broadening parameters due to the ambient pressure of He, $\textrm{H}_{2}$, and \ce{CO2} gases \citep{Tan2022-ag}, but these data are limited and not available for all of the gases in our sample. 

\begin{figure}
    \centering
    \includegraphics[width=0.5\textwidth]{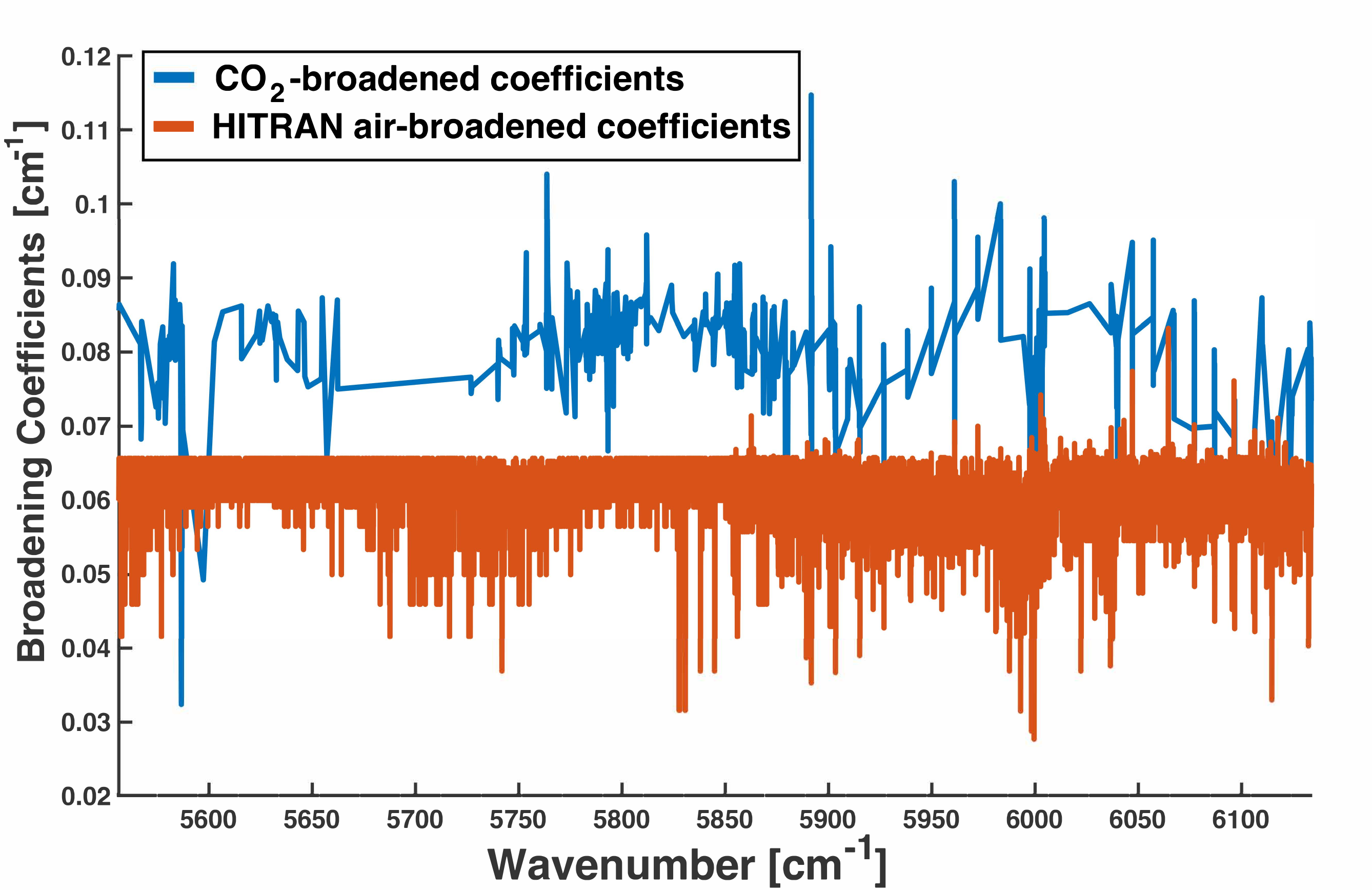}
    \caption{A comparison between \ce{CO2}-broadening coefficients \citep{Lyulin2014-kb} and air-broadening coefficients \citep{Gordon2022-mt} for \ce{CH4}. \ce{CO2}-broadened coefficients appear to differ from air-broadened coefficients by a mean factor of $\sim$1.33.}
    \label{fig:ch4_gamma}
\end{figure}

To gauge the error in radiative calculations that may arise from using air-broadening parameters in a \ce{CO2}-dominated atmosphere, we compare air-broadened coefficients (\emph{$\gamma_{air}$}) for \ce{H2O} and \ce{CH4} in the HITRAN2020 database to \ce{CO2}-broadening parameters measured and calculated by \cite{Regalia2019-tn} and \cite{Lyulin2014-kb} respectively. As shown in Figure \ref{fig:h2o_gamma} above, \ce{H2O} broadening coefficients in \ce{CO2} measured by \cite{Regalia2019-tn} are $\sim$1.6 times larger than \ce{H2O} broadening coefficients in air from the HITRAN2020 database, similar to the mean factor of 1.58 observed in \cite{Deichuli2022-sp}. Comparing the \ce{CO2}-broadening coefficients for \ce{CH4} produced by \cite{Lyulin2014-kb} to air-broadened coefficients in the HITRAN2020 database, the mean factor appears to be $\sim$1.33 as shown in Figure \ref{fig:ch4_gamma} above. 

As a test, we ran PCM LBL using a 1-bar atmosphere composed of \ce{CO2}, \ce{H2O}, and 1~ppmv of \ce{CH4} with two sets of broadening coefficients: 1) air-broadened coefficients as in the HITRAN2020 database and 2) an estimation of \ce{CO2}-broadened coefficients using the scaling factors of 1.58 and 1.33 for \ce{H2O} and \ce{CH4} respectively. The difference in OLR and surface temperature was found to be small, the latter increasing by only $\sim$0.8 K. 

\begin{figure}
    \centering
    \includegraphics[width=0.5\textwidth]{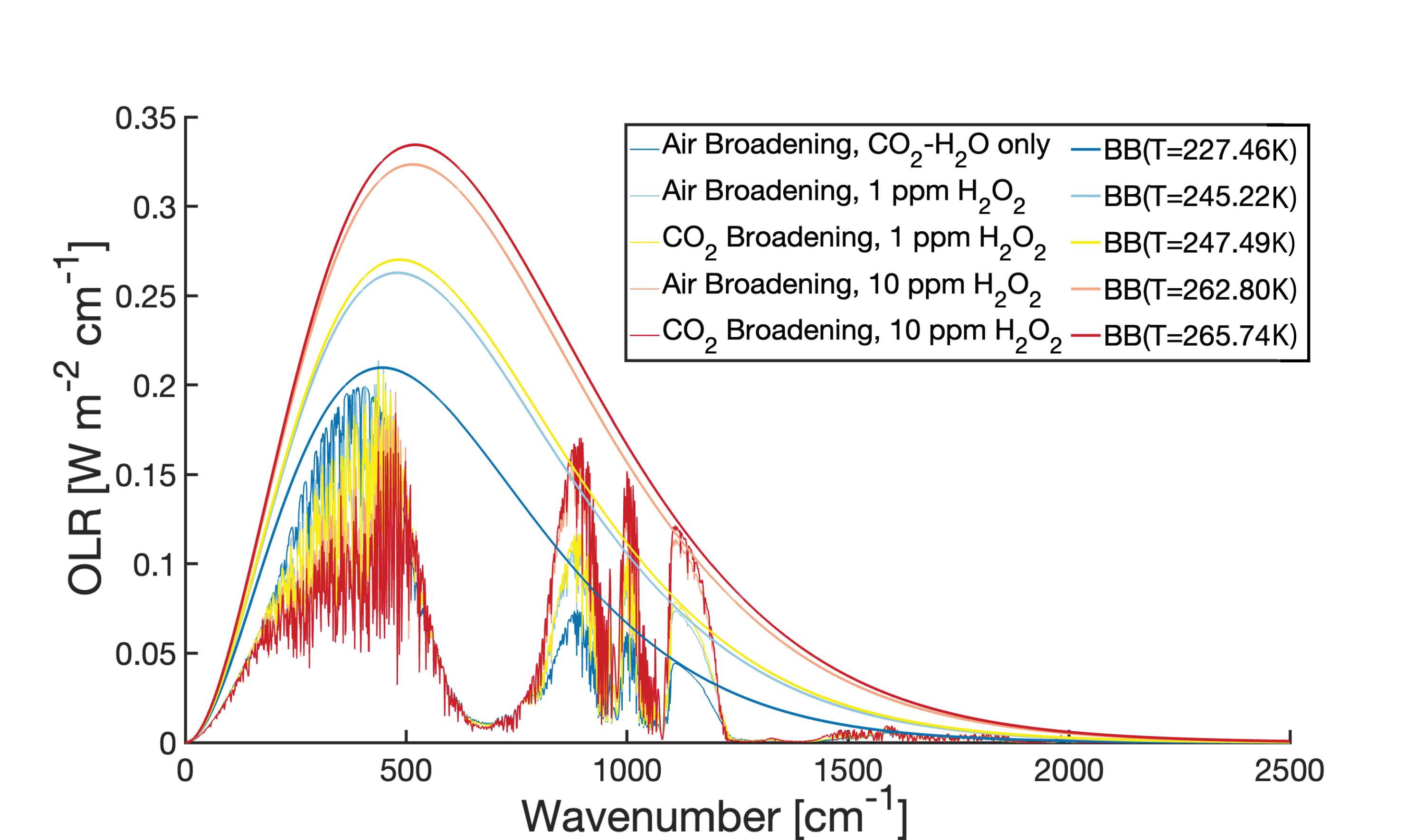}
    \caption{Comparison of OLR from a 1 bar atmosphere of \ce{CO2}, \ce{H2O}, and 1 or 10~ppmv of \ce{H2O2} given air-broadening coefficients (HITRAN data) and a reasonable estimate of the upper bound of \ce{CO2}-broadening coefficients. Differences in surface temperature between the two scenarios are on the order of $\sim$1-3K.}
    \label{fig:h2o2_gamma}
\end{figure}

Based on the scaling factors for \ce{H2O} and \ce{CH4}, we assume that doubling air-broadened parameters is a reasonable estimate of the upper bound of \ce{CO2}-broadening coefficients. To test the impact of doubled broadening parameters on an effective greenhouse gas in our sample, we additionally ran PCM LBL with each set of broadening parameters for a 1 bar atmosphere of \ce{CO2}, \ce{H2O}, and 0, 1, or 10~ppmv of \ce{H2O2}. OLR as a function of wavenumber for each of these scenarios are shown in Figure \ref{fig:h2o2_gamma}. Overlaid on each OLR plot is a blackbody curve of the same color based on the surface temperature output by the model. Doubling the air-broadening coefficients of \ce{H2O2} to estimate the \ce{CO2}-broadening coefficients increased the surface temperature $\sim$2.2~K with 1~ppmv of \ce{H2O2} and $\sim$3~K with 10~ppmv of \ce{H2O2}. 

\subsection{TIPS Ratio Approximation}

\begin{figure}
    \centering
    \includegraphics[width=0.5\textwidth]{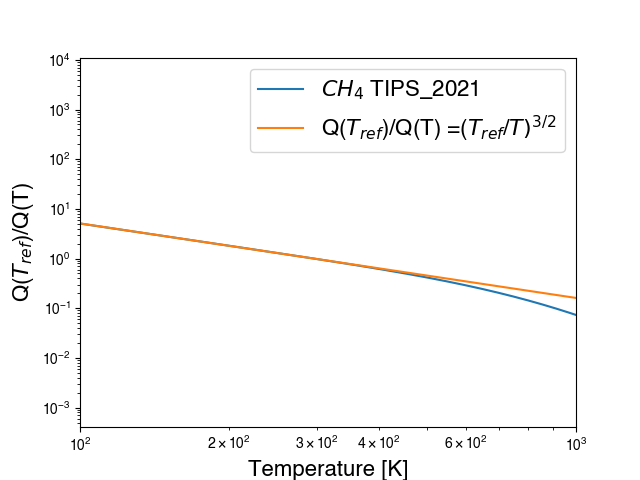}
    \caption{Comparison between TIPS and \eqref{eq:Tapprox} for the first isotopologue of \ce{CH4} between 100 K and 1000 K. \eqref{eq:Tapprox} does not begin to significantly diverse from calculated TIPS ratios until $\sim$400 K, well above the temperatures of the atmosphere of early Mars.}
    \label{fig:tips1}
\end{figure}

The Total Internal Partition Sum (TIPS) is used to determine the population of molecules as a function of quantum state. These population factors determine the strength of vibrational-rotational transitions as seen in Equation \ref{eq:linestrength} \citep{Gamache2021-wb}. For \ce{CH4}, \ce{CO}, \ce{H2O}, \ce{H2S}, \ce{HBr}, \ce{N2O}, \ce{NH3}, \ce{NO2}, \ce{O3}, \ce{HCN}, and \ce{H2CO}, we used a simple formula to approximate the TIPS ratio up to 1000 K, well above the reasonable set of temperatures we can expect in the early Martian atmosphere
\begin{equation}
Q(T_{ref})/Q(T) = (T_{ref}/T)^{1.5} \label{eq:Tapprox}.
\end{equation}
Figure ~\ref{fig:tips1} shows that \eqref{eq:Tapprox} closely matches actual \ce{CH4} TIPS values between 100~K and 300~K. However, this approximation is bad at representing TIPS values of \ce{CO2}, \ce{SO2}, \ce{H2O2}, \ce{C2H6}, \ce{C2H4}, \ce{HNO3}, and \ce{OCS} between 100~K and 1000~K. Instead, TIPS values for these gases were approximated using an 8-degree least squares polynomial fit to actual TIPS data between 100~K and 1000~K, with $r^2$ values ranging between 0.968 and 0.983. {Values used for these polynomials are given in Table~\ref{tab:poly_coeffs}.}

\section{Results}

\subsection{Simulations of All Gases at 1 bar \ce{CO2}}

\begin{table}[]
    \centering
    \begin{tabular}{c|c}
    Gas Species & GWP [$\Delta$T (K)] \\
    \hline
       \ce{H2O2}  &  47.97 \\
       \ce{HNO3}  & 42.51 \\
       \ce{SO2} & 17.67 \\
       \ce{C2H4} & 13.00 \\
       \ce{NH3} & 11.07 \\
       \ce{O3} & 6.12 \\
       \ce{OCS} & 4.89 \\
       \ce{C2H6} & 2.97 \\
       \ce{N2O} & 1.56 \\
       \ce{NO2} & 1.28 \\
       \ce{H2S} & 0.98 \\
       \ce{CH4} & 0.52 \\
       \ce{H2CO} & 0.34 \\
       \ce{HCN} & 0.19 \\
       \ce{HBr} & 0.09 \\
       \ce{CO} & 0.04 \\
    \hline
    \end{tabular}
    \caption{The difference in surface temperature between a \ce{CO2}-\ce{H2O}-only atmosphere and \ce{CO2}-\ce{H2O} atmosphere with 10~ppmv of each gas.}
    \label{tab:gwp}
\end{table}

PCM LBL produces surface temperatures of $\sim$225-227 K given a 1 bar clear-sky atmosphere of \ce{CO2} and \ce{H2O} under early Martian conditions, consistent with other modern radiative-convective models \citep{Halevy2014-jc,Ramirez2014-vt,Wordsworth2010-wu}. After reaching radiative-convective equilibrium in a pure \ce{CO2}-\ce{H2O} atmosphere, we then add 0.001 to 500~ppmv of each gas species\footnote{{Specific values used in the grid were: [0.001, 0.005, 0.01, 0.025, 0.05, 0.075, 0.1, 0.5, 1, 5, 10, 25, 50, 75, 100, 150, 200, 250, 350, 500]~ppm.}} and again ran the model to equilibrium. Table \ref{tab:gwp} shows the difference in surface temperature between a \ce{CO2}-\ce{H2O}-only atmosphere and an atmosphere after adding 10~ppmv of each gas. We define this quantity as the global warming potential, or GWP. 

 We observe roughly three regimes of GWP in our sample: 1) At 10~ppmv, strong absorbers such as \ce{H2O2} and \ce{HNO3} have GWPs above 40 K; 2) Moderate absorbers such as \ce{SO2}, \ce{C2H4}, \ce{NH3} have GWPs ranging from $\sim$11-18 K; 3) and weak absorbers have GWPs ranging from $\sim$0-6 K. 
 
\begin{figure*} 
    \centering
    \includegraphics[width=\textwidth]{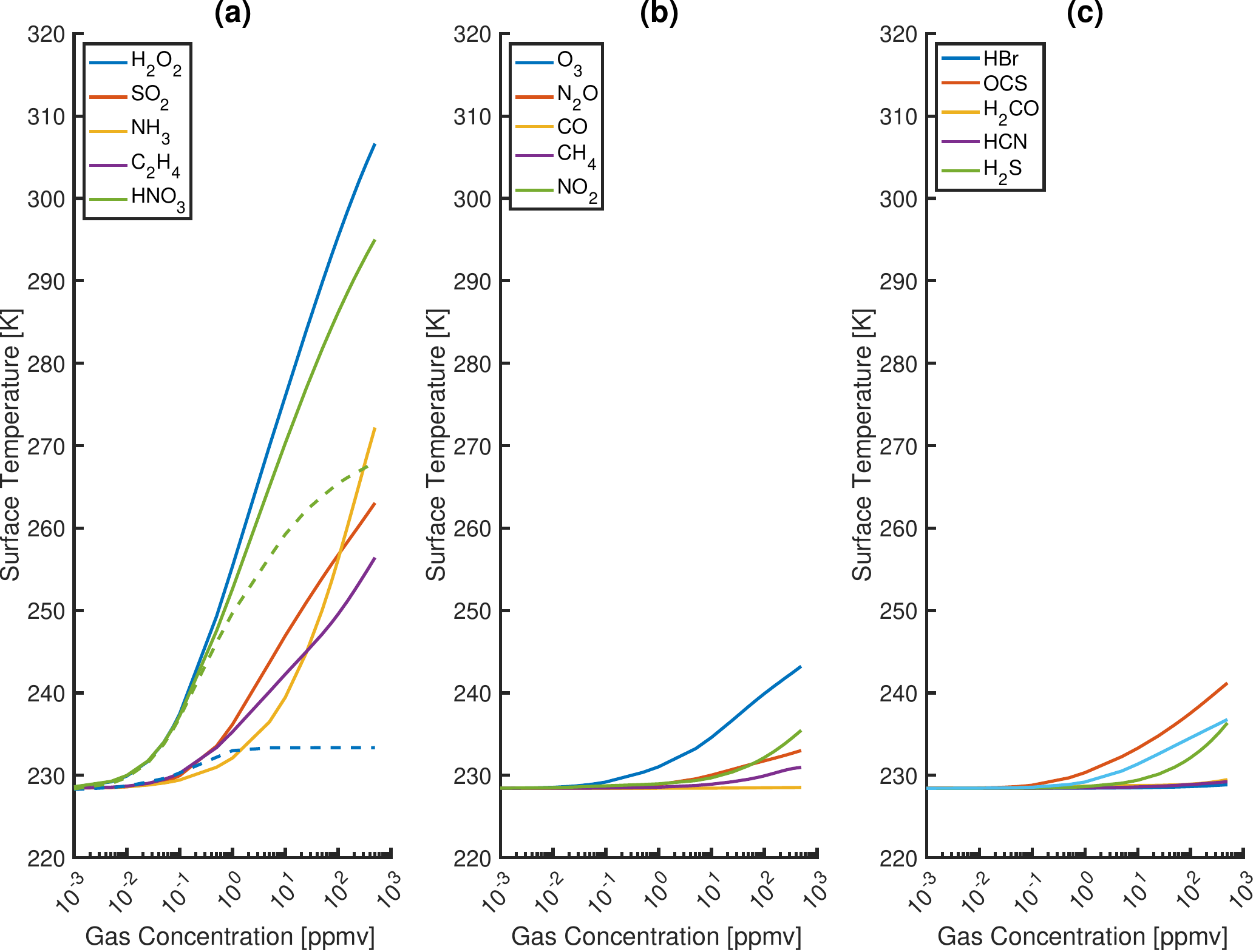}
    \caption{Surface temperature [K] vs. concentration [ppmv] for each gas in our sample, ranging from 0.001 ppmv to 500 ppmv. Strong and moderate absorbers are plotted on the left panel while weak absorbers are split between the middle and right panels for clarity. {Dashed lines show warming when gas species concentration is limited by saturation.} \ce{H2O2} and \ce{HNO3} are the most effective greenhouse gases in our sample, {and the most strongly affected by condensation.}} 
    \label{fig:Tsurf_conc}
\end{figure*}

Figure \ref{fig:Tsurf_conc} shows surface temperature [K] as a function of concentration [ppmv] for each gas species. Strong and moderate absorbers are shown on the left panel while the remainder are split between the middle and right panels. \ce{H2O2} and \ce{HNO3} are the strongest absorbers across all tested concentrations and are able to warm the early Martian surface by multiple degrees, even at extremely low concentrations of 0.001 ppmv and 0.01 ppmv. At 10~ppmv and above, \ce{H2O2} and \ce{HNO3} each push the surface temperature toward $>$273 K in our 1 bar atmosphere, demonstrating their uniquely high radiative potential. {However, as the blue and green dashed lines indicate, these concentrations can only be achieved if the gases are supersaturated in the martian atmosphere. Condensation most severely limits \ce{H2O2} concentrations, causing greenhouse warming to be limited to a little over 230~K in that case.}

Moderate absorbers such as \ce{SO2}, \ce{C2H4}, and \ce{NH3} begin to meaningfully warm the surface at concentrations ranging from 0.5 to 1~ppmv, with GWPS rising from $\sim$4-7 K at 1~ppmv to $\sim$20-28 K at 100~ppmv. The GWPs of the remaining gases at 10~ppmv range from 0.04 K to 6.1 K. While \ce{CO}, \ce{HBr}, \ce{HCN},and \ce{H2CO} have the smallest radiative effect across concentration parameter space, \ce{O3}, \ce{OCS}, and \ce{C2H6} warm the surface by $\sim$6.1 K, $\sim$4.9 K, and $\sim$3 K respectively. These gases could be important for warming early Mars, but higher concentrations are required to have a similar radiative impact to that of \ce{H2O2}, \ce{HNO3}, and to a lesser extent \ce{SO2}, \ce{C2H4}, and \ce{NH3}. While we do not focus on \ce{O3}, \ce{OCS}, and \ce{C2H6} in the following discussion, future studies could also be conducted to evaluate their atmospheric sources and sinks on early Mars.
\subsection{Simulations of Select Gases at 0.5 and 2 bars}

From our radiative calculations of all gases at 1 bar, we narrow our focus to the five strongest absorbers in our sample: \ce{H2O2}, \ce{HNO3}, \ce{SO2}, \ce{C2H4}, and \ce{NH3}. We simulate a clear-sky atmosphere of \ce{CO2} and \ce{H2O} under early Martian conditions at 0.5 bars and 2 bars  with the same concentrations as before, adding our strong and moderate absorbers. Given the total atmospheric pressure on early Mars is not well constrained, it is important to consider how the radiative effects of our strong and moderate absorbers change given different maximum surface pressures. Because the atmosphere collapses onto the surface completely at high enough \ce{CO2} pressures \citep{Kasting1991-xx,Forget2013-zd,soto2015martian} and there is not evidence for substantial carbon burial \citep{Jakosky2019-rf,Hu2015-fc,Wordsworth2016-ua}, 2 bars is chosen to be an upper bound on total atmospheric pressure. Atmospheric collapse into ice caps can also occur at low total atmospheric pressures depending on the obliquity of the planet, so 0.5 bars is chosen a lower bound on total atmospheric pressure. 

\begin{figure*}
    \centering
    \includegraphics[width=\textwidth]{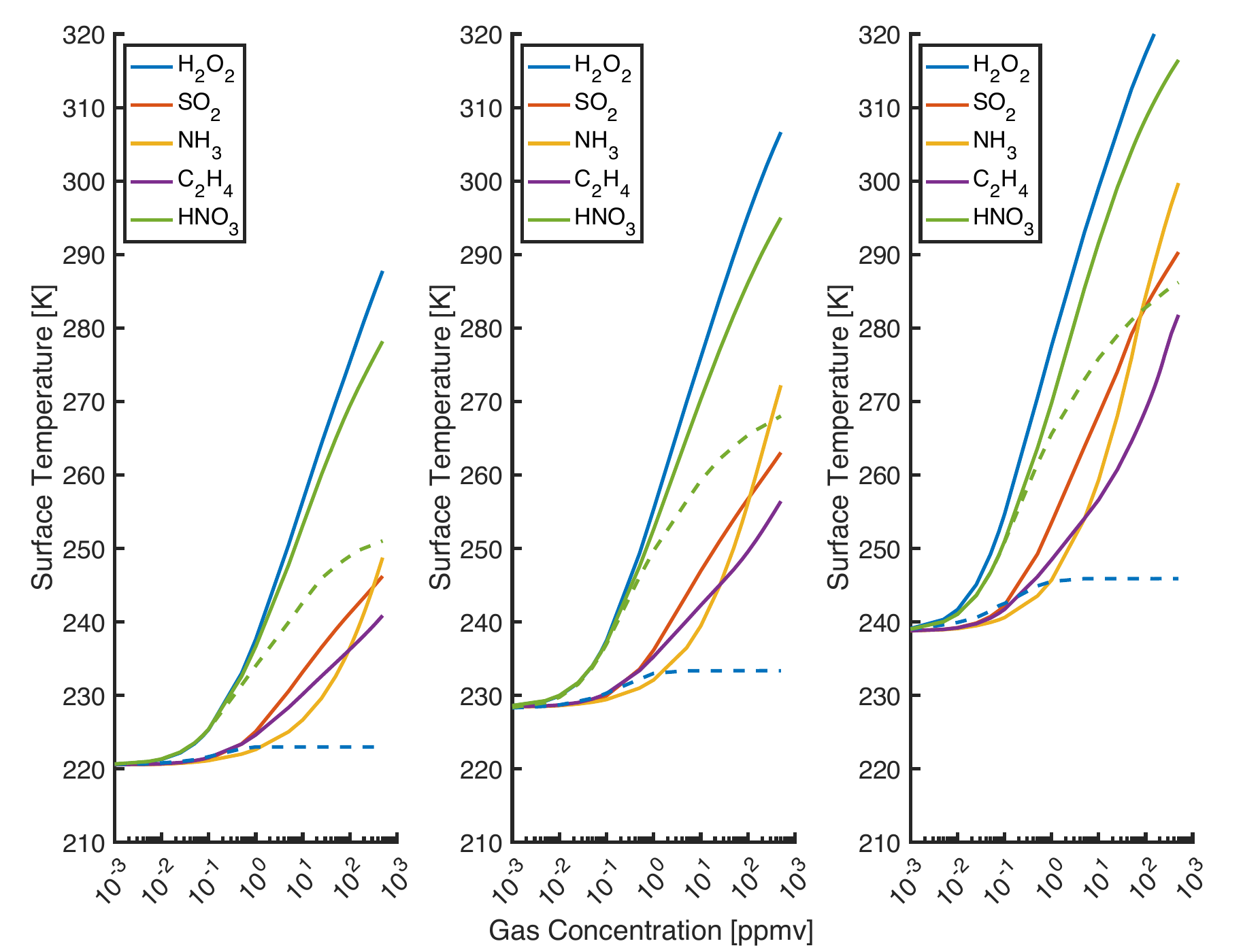}
    \caption{Surface temperature [K] vs. gas concentration [ppmv] of \ce{H2O2}, \ce{HNO3}, \ce{SO2}, \ce{C2H4}, and \ce{NH3} at 0.5, 1, and 2 bars of total atmospheric pressure. As maximum surface pressure increases, so does the greenhouse effect induced by each gas.}
    \label{fig:p_multi}
\end{figure*}

The results of these radiative calculations are shown in Figure \ref{fig:p_multi}. At 0.5 bars, the differences between strong and moderate absorbers is more distinct. \ce{H2O2} and \ce{HNO3} warm the surface $\sim$10 K at 1~ppmv, $\sim$20 K at 10~ppmv, and $\sim$40 K at 100~ppmv. Each is capable of increasing surface temperature to $>$273~K alone, but only at the extreme end of our concentration parameter space (350-500~ppmv). In contrast, \ce{SO2}, \ce{C2H4}, and \ce{NH3} only warm the surface $\sim$1-3 K at 1~ppmv, $\sim$5-10 K at 10~ppmv, and $\sim$10-15 K at 100~ppmv. At 2 bars, there is less distinction between strong and moderate absorbers across the surface temperature-concentration parameter space. \ce{H2O2} and \ce{HNO3} push the surface temperature to $>$273 K between 1 and 10~ppmv, while \ce{SO2} and \ce{NH3} do so close to 100~ppmv, and \ce{C2H4} does so at around 500~ppmv. 

\begin{figure}
    \centering
    \includegraphics[width=0.5\textwidth]{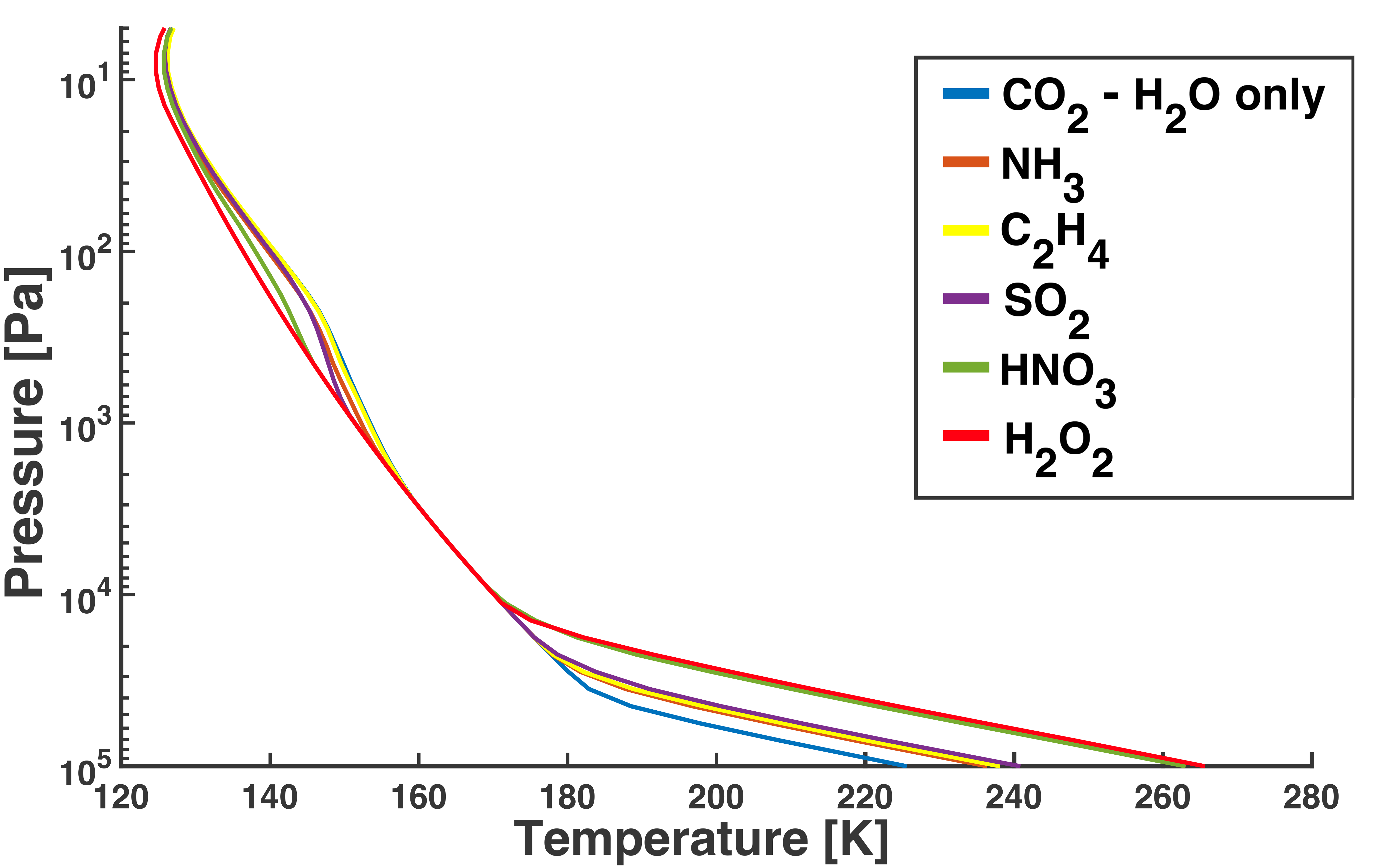}
    \caption{Temperature-pressure profiles of 1 bar \ce{CO2}-\ce{H2O} atmospheres with the addition of 10~ppmv of \ce{H2O2} (red), \ce{HNO3} (green), \ce{SO2} (purple), \ce{C2H4} (yellow), and \ce{NH3} (orange). 
    A pure \ce{CO2}-\ce{H2O} profile is also shown as a comparison (blue). }
    \label{fig:TvsZ}
\end{figure}

Figure \ref{fig:TvsZ} shows the temperature as a function of altitude from the surface to the top-of-atmosphere at 1 Pa for a 1 bar \ce{CO2}-\ce{H2O} atmosphere with the addition of 10~ppmv of \ce{H2O2}, \ce{HNO3}, \ce{SO2}, \ce{C2H4}, and \ce{NH3}. Each temperature-pressure profile shows temperature following a moist adiabat in the lower troposphere from the surface until about 0.5 bars, followed by \ce{CO2} condensation in the middle atmosphere between $\sim$0.01 and 0.1 bars where each profile converges. Above 0.01 bars, the profiles diverge as the upper atmosphere reaches radiative equilibrium. In this case, \ce{H2O2} and \ce{HNO3} warm the surface 15-30 K more than \ce{SO2}, \ce{C2H4}, and \ce{NH3}. 

\ce{H2O2}, \ce{HNO3}, \ce{SO2}, \ce{C2H4}, and \ce{NH3} are effective greenhouse gases under early Martian conditions because they absorb in atmospheric window regions around 400~\invcm and 1000~\invcm. \ce{CO2} is opaque due to the $\nu_2$ 667~\invcm (15 $\mu$m) band \citep[e.g.,][]{Goody1995-pp,Wordsworth_2024} and CIA due to induced dipole effects around 1-250 \invcm \citep{Gruszka1998-vk} and dimer effects from 1200-1500 \invcm\citep{Baranov2004-xp}. While \ce{CO2} CIA absorption effects are taken into account, \ce{CO2} CIA absorption bands are not included in absorption cross-section files. Water vapor is a strong absorber at low wavenumbers and at its $\nu_2$ band at 1600~\invcm.

\begin{figure*}
    \centering
        \includegraphics[width=\textwidth]{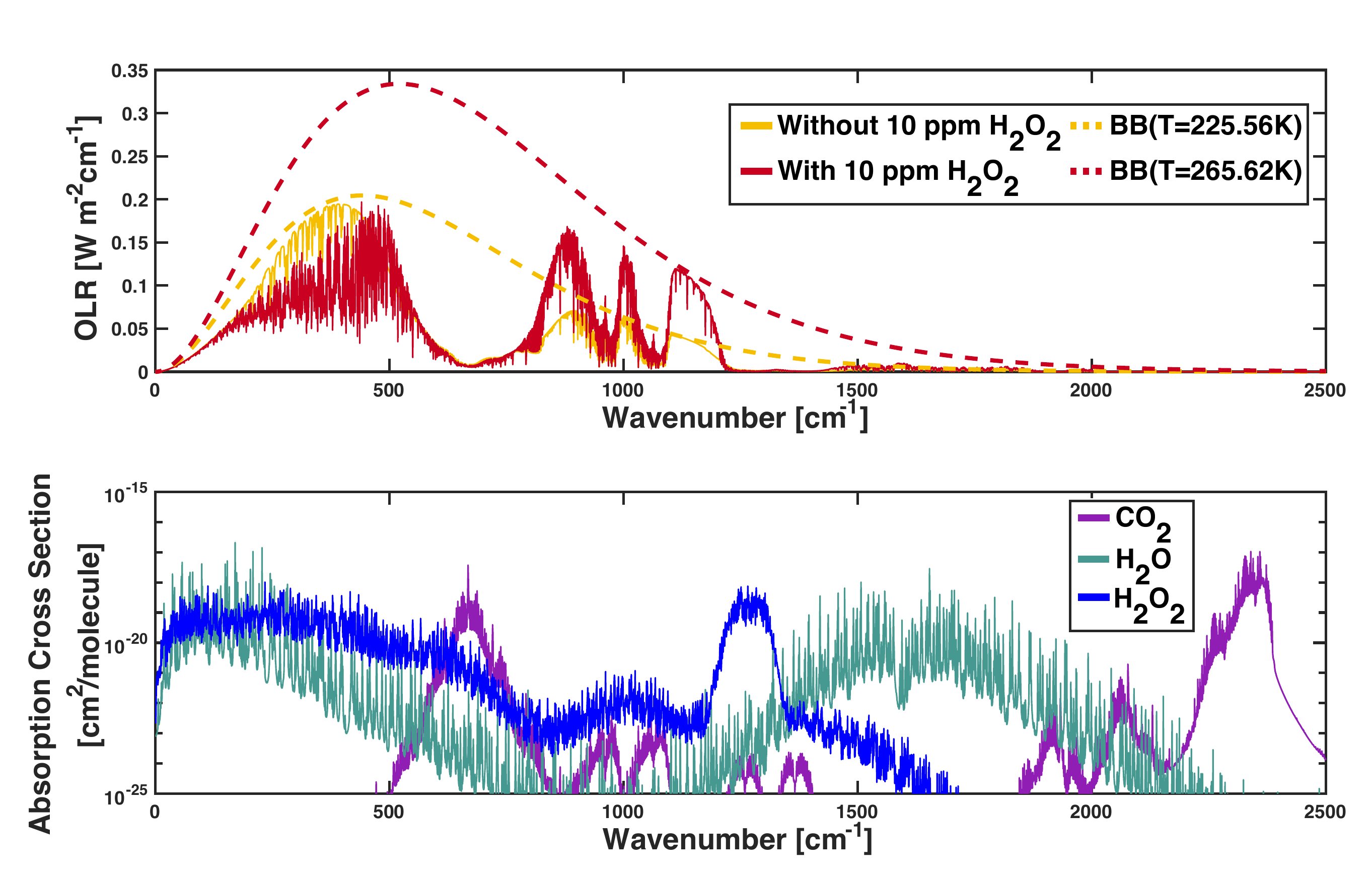}
    \caption{Upper panel: Outgoing planetary radiation as a function of wavenumber for a 1 bar atmosphere of \ce{CO2} and \ce{H2O} with (red) and without (yellow) 10~ppmv of \ce{H2O2}. The dashed colored curves show blackbody radiation, of which the temperatures are indicated by BB(T). Lower panel: Absorption cross sections of \ce{CO2} (purple), \ce{H2O} (teal), and \ce{H2O2} (blue) at 1 bar, as functions of wavenumber.}
    \label{fig:H2O2}
\end{figure*}

Shown in Figure \ref{fig:H2O2}, \ce{H2O2} absorbs radiation at wavenumbers between 200 and 500 \invcm, near the peak of the blackbody spectrum for a temperature of 265~K. \ce{H2O2} also has a strong absorption band centered around 1250 \invcm. Absorption in these atmospheric window regions by \ce{H2O2} where there is not significant interference by \ce{CO2} and \ce{H2O} absorption bands pushes the surface temperature toward 273~K, making it an effective greenhouse gas under early Martian conditions.

\begin{figure*}
    \centering
    \includegraphics[width=\textwidth]{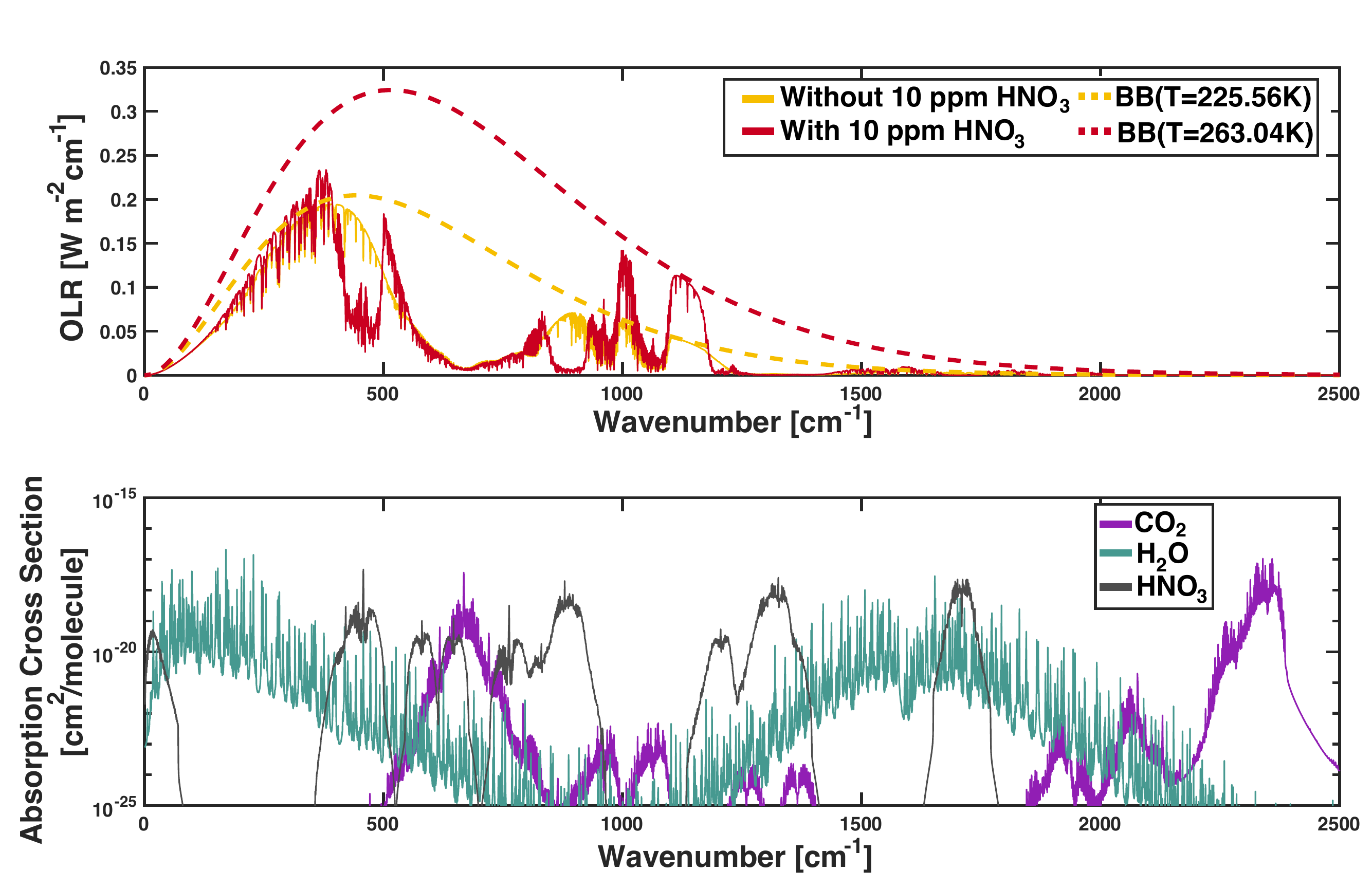}
    \caption{Upper panel: Outgoing planetary radiation as a function of wavenumber for a 1 bar atmosphere of \ce{CO2} and \ce{H2O} with (red) and without (yellow) 10~ppmv of \ce{HNO3}. The dashed colored curves show blackbody radiation, of which the temperatures are indicated by BB(T). Lower panel: Absorption cross sections of \ce{CO2} (purple), \ce{H2O} (teal), and \ce{HNO3} (grey) at 1 bar, as functions of wavenumber.}
    \label{fig:HNO3}
\end{figure*}

\begin{figure*}
    \centering
    \includegraphics[width=\textwidth]{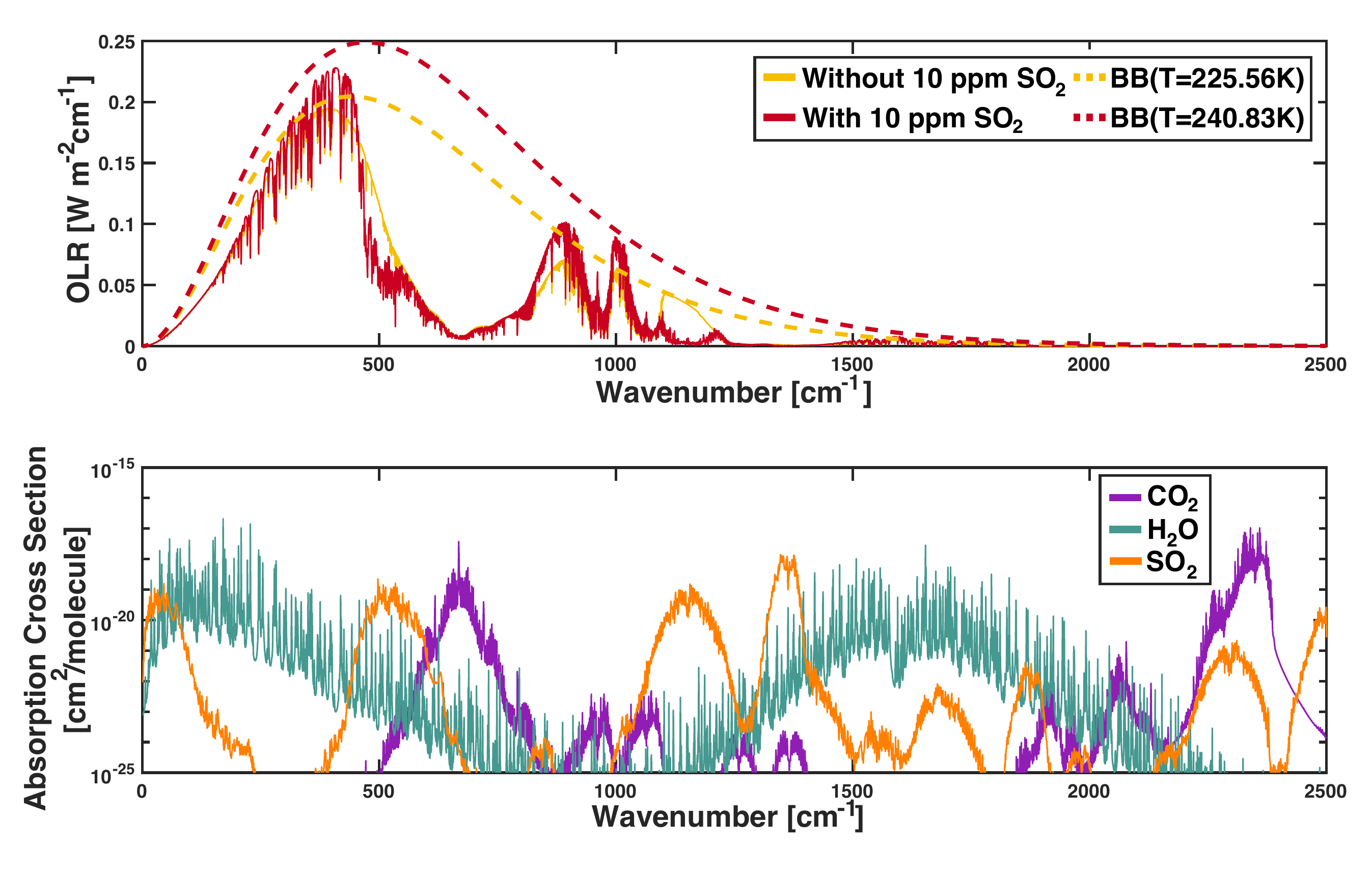}
    \caption{Upper panel: Outgoing planetary radiation as a function of wavenumber for a 1 bar atmosphere of \ce{CO2} and \ce{H2O} with (red) and without (yellow) 10~ppmv of \ce{SO2}. The dashed colored curves show blackbody radiation, of which the temperatures are indicated by BB(T). Lower panel: Absorption cross sections of \ce{CO2} (purple), \ce{H2O} (teal), and \ce{SO2} (orange) at 1 bar, as functions of wavenumber.}
    \label{fig:SO2}
\end{figure*}

\begin{figure*}
    \centering
    \includegraphics[width=\textwidth]{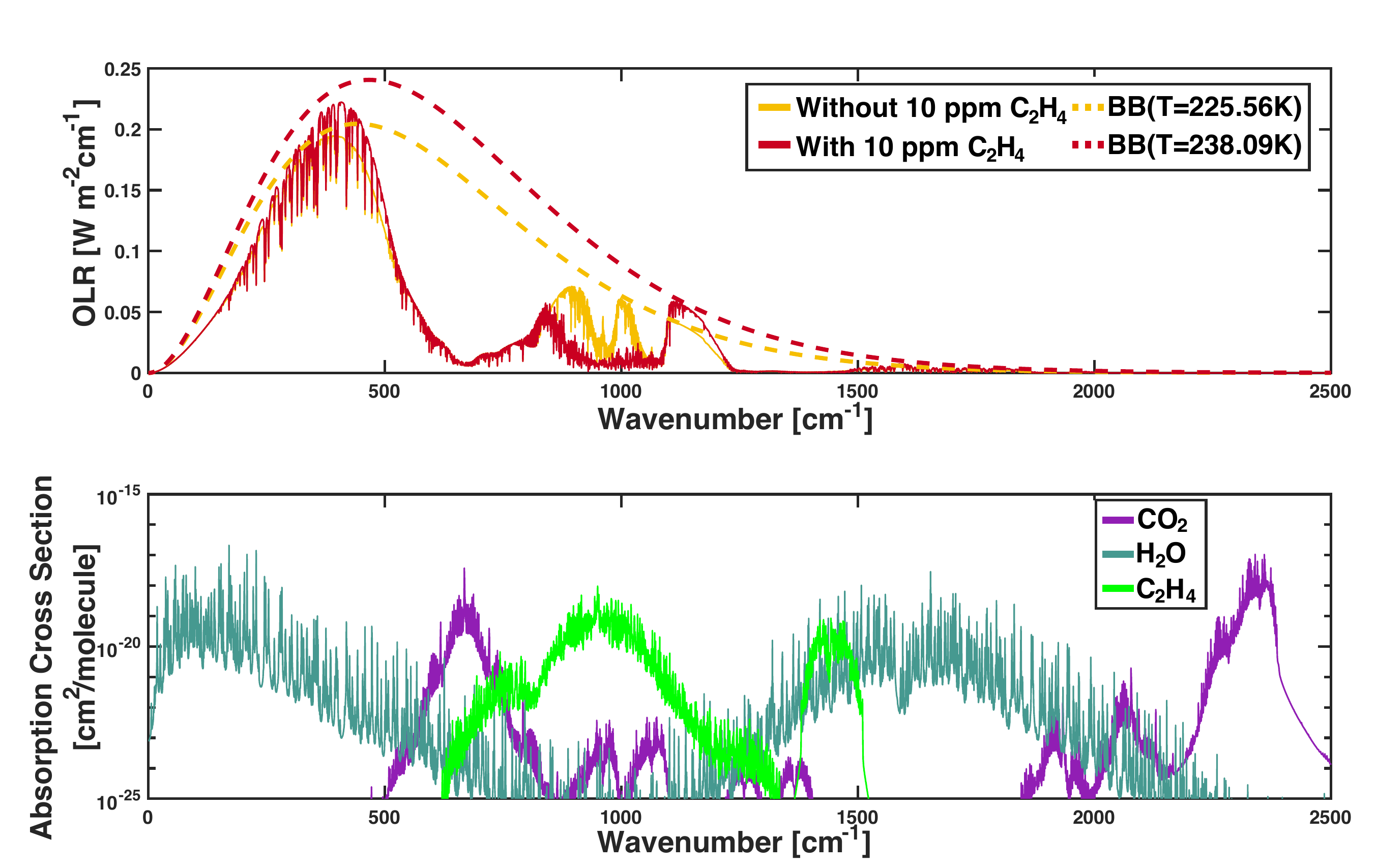}
    \caption{Upper panel: Outgoing planetary radiation as a function of wavenumber for a 1 bar atmosphere of \ce{CO2} and \ce{H2O} with (red) and without (yellow) 10~ppmv of \ce{C2H4}. The dashed colored curves show blackbody radiation cu, of which the temperatures are indicated by BB(T). Lower panel: Absorption cross sections of \ce{CO2} (purple), \ce{H2O} (teal), and \ce{C2H4} (green) at 1 bar, as functions of wavenumber.}
    \label{fig:C2H4}
\end{figure*}

\begin{figure*}
    \centering
    \includegraphics[width=\textwidth]{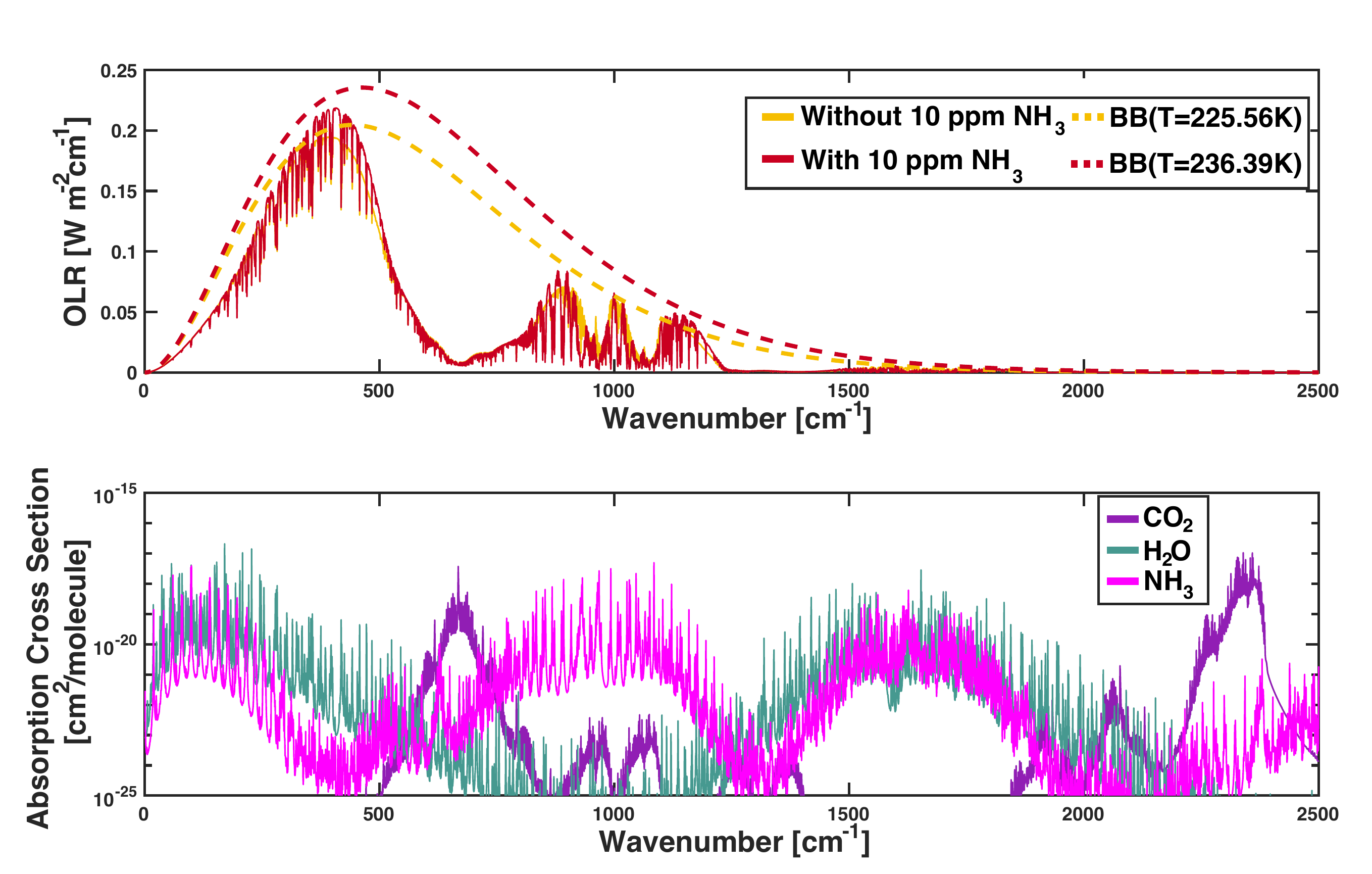}
    \caption{Upper panel: Outgoing planetary radiation as a function of wavenumber for a 1 bar atmosphere of \ce{CO2} and \ce{H2O} with (red) and without (yellow) 10~ppmv of \ce{NH3}. The dashed colored curves show blackbody radiation, of which the temperatures are indicated by BB(T). Lower panel: Absorption cross sections of \ce{CO2} (purple), \ce{H2O} (teal), and \ce{NH3} (pink) at 1 bar, as functions of wavenumber.}
    \label{fig:NH3}
\end{figure*}

\ce{HNO3} is a very effective greenhouse gas. Shown in Figure \ref{fig:HNO3}, \ce{HNO3} has a strong absorption band centered just before 500 \invcm near the peak of the blackbody spectrum at $\sim$260 K. This absorption band only partially intersects \ce{H2O} absorption and avoids intersecting with the $\nu_2$ 667 \invcm \ce{CO2} absorption band.
\ce{HNO3} has another set of absorption bands centered around 800-1000 \invcm, a region where there is neither strong \ce{CO2} nor \ce{H2O} absorption. This absorption band plays a significant role in making \ce{HNO3} an effective minor greenhouse gas under early Martian conditions.

\ce{SO2} is a moderately effective greenhouse gas. Shown in Figure \ref{fig:SO2}, \ce{SO2} has an absorption band centered around 500 \invcm, just to the right of the peak of the blackbody spectrum at 240 K. However, this absorption band intersects with the $\nu_2$ 667 \invcm  \ce{CO2} absorption band. Because the atmosphere is already opaque at these wavelengths due to \ce{CO2} absorption, \ce{SO2} absorption has less of an effect on the surface temperature. While an \ce{SO2} absorption band centered around 1100 \invcm absorbs in an atmospheric window region and reduces OLR, absorption bands in the 1200-1500 \invcm region partially intersect with \ce{CO2} CIA absorption bands. This reduces \ce{SO2}'s effectiveness as a minor greenhouse gas compared to \ce{HNO3}. 

\ce{C2H4} is a moderately effective greenhouse gas. Shown in Figure \ref{fig:C2H4}, \ce{C2H4} has two main absorption bands centered around 1000 \invcm and 1400 \invcm. The 1000 \invcm band absorbs in an atmospheric window region and is only intersected by weak pair of \ce{CO2} absorption bands centered about 1000 \invcm, avoiding intersecting with the $\nu_2$ 667 \invcm  \ce{CO2} absorption band and \ce{CO2} CIA absorption bands from 1200-1500 \invcm . However, the \ce{C2H4} band centered around 1150 \invcm intersects with absorption by \ce{H2O} and is too far from the peak of the blackbody spectrum at 240 K to have an appreciable effect on surface temperature.

Finally, \ce{NH3} is a moderately effective greenhouse gas. Shown in Figure \ref{fig:NH3}, \ce{NH3} has absorption bands centered around 100 \invcm, 1000 \invcm, and 1600 \invcm. The absorption bands around 100 \invcm and 1600 \invcm are intersected by \ce{H2O} absorption and do not contribute to reductions in OLR, and the latter is far from the peak of the blackbody spectrum at $\sim$240 K. While the \ce{NH3} absorption band centered around 1000 \invcm is only intersected by weak \ce{CO2} absorption bands, it is very broad and less effective at absorption. Therefore, the reductions in OLR that arise from the portion of the \ce{NH3} absorption band around 1000 \invcm that is not intersected by \ce{CO2} or \ce{H2O} absorption are only modest.

\section{Discussion}

Our comprehensive survey deliberately focuses on the warming potential of all known greenhouse gases on early Mars to provide a database of greenhouse warming. While our detailed calculations show that numerous gases, including \ce{H2O2}, \ce{HNO3}, \ce{NH3}, \ce{SO2}, and \ce{C2H4}, could act to warm the early Martian surface, it is important to couple climate modeling to atmospheric chemistry modeling and study sources and sinks to evaluate how much each greenhouse gas might accumulate in Mars' early atmosphere. Although this work is outside the scope of this paper,  we qualitatively consider some of the wider context here.

{When supersaturated concentrations are permitted, } we find that \ce{H2O2} is the most effective minor greenhouse gas of those modeled, supporting the climate results of \cite{Ito2020-me}. While \ce{H2O2} is the product of reactions between \ce{HOx} species in a wet and oxidized atmosphere that may have existed at least episodically on early Mars \citep{Hurowitz2017-gd,Wordsworth2021-qg}, \ce{H2O2} {condenses at low concentrations}, is destroyed easily by photodissociation and rainout, as it is extremely soluble \citep{Zahnle2008-iy}. Dry deposition is another potentially important removal mechanism of \ce{H2O2}, where the turbulence, solubility and reactivity of \ce{H2O2}, and the compositions and oxidation states of the surface affect the amount taken up at the surface.

\ce{HNO3} is the second most effective minor greenhouse gas of those modeled. While the Mars Science Laboratory aboard Curiosity recently discovered nitrates in Gale Crater \citep[e.g.,][]{Stern2015-cu}, \ce{HNO3} has not been extensively studied as a potential greenhouse gas on early Mars. \cite{Adams2021-yn} suggest possible mechanisms of nitrate deposition on early Mars due to both the formation of $\textrm{HNO}_{x}$ from lightning-induced NO, {and splitting of \ce{N2} by solar energetic particles created by coronal mass ejection events on the presumably more active young Sun.}  \cite{Wong2017-dn} similarly propose that the \ce{CO2}-\ce{N2} dominated atmosphere of the Hadean Earth would, when heated by lightning, produce NO. NO would then react with water vapor to produce produce acids such as HNO, $\textrm{HNO}_{2}$, and \ce{HNO3} \citep{Wong2017-dn}. \cite{Adams2021-yn} studies the photochemical production and precipitation of $\textrm{HNO}_{x}$ in an early Martian environment through a photochemistry and transport model, finding that \ce{HNO3} is formed predominantly through the reaction of NO and $\textrm{HO}_{2}$ as in the mechanism proposed by \cite{Wong2017-dn}. The model used in \cite{Adams2021-yn} also produces NO fluxes on early Mars similar to those calculated by \cite{Wong2017-dn} for early Earth, suggesting the production of \ce{HNO3} could occur on early Mars. However, as for \ce{H2O2}, the greatest challenge for \ce{HNO3} is that it is {relatively easy to condense and} extremely soluble in water, making it extremely hard to build up in an atmosphere with an active hydrological cycle. 

We find that \ce{SO2} is one of the most effective minor greenhouse gases of those modeled. An \ce{SO2} greenhouse on early Mars produced by extensive volcanism has been studied previously as a solution to the early Mars climate problem \citep[e.g.,][]{Halevy2014-jc}, but \ce{SO2} forms \ce{H2SO4} and elemental-sulfur aerosols in weeks to months followings volcanic eruptions, causing cooling. The anti-greenhouse effect of these aerosols likely outweighs the greenhouse effect from sulfur gas in the long term \citep{Tian2010-pb,kerber2015sulfur}. \ce{SO2} is also rather soluble in water and easily removed from the atmosphere by rainout \citep{Johnson2009-as}.

Our radiative calculations also find that \ce{NH3} acts as a potent greenhouse gas. Reducing gases such as \ce{CH4} and \ce{NH3} have been proposed as a solution to the early Mars climate problem, with \cite{Sagan1977-ae} suggesting that a hydrogen-dominated atmosphere or copious amounts of \ce{NH3} could account for evidence of fluvial activity on the surface. Furthermore, \cite{Kasting1992-ko} finds that 500~ppmv of \ce{NH3} in a 4-5 bar \ce{CO2} atmosphere could raise the surface temperature above 273 K. However, \ce{NH3} is readily photolyzed by UV radiation on short time scales and there are not any significant plausible Martian sources. Appreciable concentrations of \ce{NH3} could exist if there were a large surface or subsurface source and it was shielded from photolysis by an high-altitude organic haze \citep{Sagan1997-bf,Wolf2010-lu}.

Finally, our radiative calculations show that \ce{C2H4} is a moderately effective greenhouse gas under early Martian conditions. While \ce{C2H4} photochemistry on early Mars has been considered as an eventual product of \ce{CH4} photolysis and the reaction of methyl radicals, it has not been widely studied as a potential greenhouse gas. \cite{Haqq-Misra2008-uh} investigate the radiative effects of a hazy methane greenhouse on the Archean Earth and argue that significant concentrations of \ce{CH4} and solar UV radiation lead to the formation of higher-chain hydrocarbons such as \ce{C2H6} and \ce{C2H4}, possible explanations for the Faint Young Sun paradox. Photolysis or oxidation of \ce{CH4} leads to the formation of methyl radicals that can react to eventually form \ce{C2H4}; significant concentrations of these gases are often accompanied by organic haze formation whose thickness determines its net radiative effect \citep{Haqq-Misra2008-uh}. \cite{Trainer2006-rf} consider a similar organic haze on Titan and early Earth, arguing that the irradiation of a \ce{CH4}-\ce{N2} atmosphere leads to the formation of aerosol particles that aggregate into fractal agglomerates and produce a thick haze layer. However, \cite{Trainer2006-rf} note that haze photochemistry in a bulk \ce{CO2} atmosphere likely differs from that observed on Titan. More photochemical modeling, ideally with additional laboratory data focused on haze formation in \ce{CH4}-\ce{CO2} atmospheres, is needed to determine how much \ce{C2H4} could have warmed the early climate.

\section{Conclusion}

We have performed the first general survey of the warming potential of greenhouse gases on early Mars. Our analysis has revealed several important phenomena. First, the most radiatively active gas species also tend to be those that are highly soluble and/or vulnerable to photolytic destruction. This can be explained qualitatively by the fact that molecules with a large number of vibrational and rotational states are often polar, which allows them to dissolve easily in liquid water, and also makes them easy to break apart with moderately high energy photons.

We find that both reducing (\ce{C2H4}, \ce{NH3}) and oxidizing (\ce{H2O2}, \ce{HNO3}, \ce{SO2}) minor species have the ability to cause greenhouse warming when present in high enough concentrations. This is interesting in the context of the martian geologic record, which strongly suggests that Mars has passed through both reducing and oxidizing atmospheric states in the past, with a slow trend to more oxidizing conditions over geologic time driven by loss of H to space \citep{Hurowitz2017-gd,Wordsworth2021-qg,liu2021anoxic,Jakosky2018-lj}.

This study has focused on background atmospheres dominated by \ce{CO2}, and addition of \ce{N2} to the calculations \citep[e.g., ][]{hu2022nitrogen} could be expected to have some effect on the results. Additional climatic limitations of this study are that we have neglected 3D atmospheric dynamics and cloud radiative effects. {Although such an assumption is standard in one-dimensional radiative-convective modeling, clouds can affect the radiative forcing of greenhouse gases via both scattering and absorption, with the severity of the modification  dependent on microphysics and global distribution.} This could be addressed in future with 3D GCM simulations. However, the most important next step will be to use photochemical models with realistic representations of sources and sinks, in order to evaluate the plausibility of high enough concentrations to cause greenhouse warming for any of the species modeled here.

\section{Acknowledgements}
Support for this work was provided by NSF CAREER award AST-1847120 and NASA VPL award UWSC10439. The authors would like to thank Jake Seeley for helpful comments on line-by-line radiative calculations and Iouli Gordon for insight into the HITRAN spectroscopic database and TIPS calculations.

\section{Data Availability Statement}

{Surface temperature data tables, as well as the source code for the PCM LBL model used to produce all figures in the study, are available open-source at \emph{https://zenodo.org/records/13774100} \citep{zenodo_reference}. }

\newpage
\section{Appendix}

{
\begin{sidewaystable}
\centering
\begin{tabular}{|c|c|c|c|c|c|c|c|c|c|}
\hline
Gas Type & $a_0$ & $a_1$ & $a_2$ & $a_3$ & $a_4$ & $a_5$ & $a_6$ & $a_7$ & $a_8$ \\ \hline
\ce{H2O2} & 51.906 & -0.736 & 0.0049 & -1.867$\times10^{-5}$ & 4.37$\times10^{-8}$ & -6.36$\times10^{-11}$ & 5.617$\times10^{-14}$ & -2.749$\times10^{-17}$ & 5.718$\times10^{-21}$ \\ \hline
\ce{HNO3} & 51.578 & -0.718 & 0.0047 & -1.794$\times10^{-5}$ & 4.21$\times10^{-8}$ & -6.156$\times10^{-11}$ & 5.454$\times10^{-14}$ & -2.678$\times10^{-17}$ & 5.585$\times10^{-21}$ \\ \hline
\ce{C2H6} & 55.45 & -0.769 & 0.005 & -1.918$\times10^{-5}$ & 4.474$\times10^{-8}$ & -6.501$\times10^{-11}$ & 5.727$\times10^{-14}$ & -2.798$\times10^{-17}$ & 5.813$\times10^{-21}$ \\ \hline
\ce{CO2} & 13.50 & -0.156 & 0.000968 & -3.613$\times10^{-6}$ & 8.37$\times10^{-9}$ & -1.217$\times10^{-11}$ & 1.065$\times10^{-14}$ & -5.197$\times10^{-18}$ & 1.078$\times10^{-21}$ \\ \hline
\ce{C2H4} & 36 & -0.499 & 0.0033 & -1.271$\times10^{-5}$ & 2.982$\times10^{-8}$ & -4.347$\times10^{-11}$ & 3.838$\times10^{-14}$ & -1.878$\times10^{-17}$ & 3.905$\times10^{-21}$ \\ \hline
\ce{SO2} & 36.47 & -0.502 & 0.0033 & -1.262$\times10^{-5}$ & 2.954$\times10^{-8}$ & -4.302$\times10^{-11}$ & 3.799$\times10^{-14}$ & -1.86$\times10^{-17}$ & 3.87$\times10^{-21}$ \\ \hline
\ce{OCS} & 15.99 & -0.181 & 0.0011 & -4.099$\times10^{-6}$ & 9.507$\times10^{-9}$ & -1.381$\times10^{-11}$ & 1.218$\times10^{-14}$ & -5.964$\times10^{-18}$ & 1.241$\times10^{-21}$ \\ \hline
\end{tabular}
\caption{Coefficients of the total internal partition sum polynomial $Q(T) = a_0 + a_1 T + a_2 T^2 + \ldots + a_8 T^8$ for different gases.}
\label{tab:poly_coeffs}
\end{sidewaystable}
}

\end{document}